\documentclass{article}

\usepackage{epsfig}
\usepackage{graphicx}
\usepackage{amssymb}
\usepackage{amsfonts,amsbsy}
\usepackage{psfrag}
\usepackage{amsmath}
\usepackage{tikz}
\usepackage{subfigure}
\usepackage{wasysym}

\def\e{\begin{equation}} 
\def\f{\end{equation}} 
\def\ea{\begin{eqnarray}} 
\def\fa{\end{eqnarray}} 

\def\##1{{\mbox{\textbf{#1}}}}
\def\%#1{{\mbox{\boldmath $#1$}}}

\def\=#1{{\overline{\overline{\mathsf #1}}}}

\def\/{\over}
\def\*{^{\displaystyle*}}

\def\.{\cdot}
\def\x{\times}
\def\:{\over}
\def\oo{\infty}

\def\ra{\rightarrow}

\def\le{\left(}
\def\ri{\right)}
\def\l#1{\label{eq:#1}}
\def\r#1{(\ref{eq:#1})}
\def\am{\left(\begin{array}{c}}
\def\amm{\left(\begin{array}{cc}}
\def\ammm{\left(\begin{array}{ccc}}
\def\ammmm{\left(\begin{array}{cccc}}
\def\a{\end{array}\right)}

\def\add{\left|\begin{array}{cc}}
\def\addd{\left|\begin{array}{ccc}}
\def\adddd{\left|\begin{array}{cccc}}
\def\ad{\end{array}\right|}

\def\A{\alpha}
\def\B{\beta}

\def\E{\epsilon}

\def\h{\eta}

\def\M{\mu}
\def\o{\omega}

\def\tr{{\rm tr }}

\def\det{{\rm det}}

\begin{document}

\title{Electromagnetic Boundary Conditions Defined by Reflection Properties of Eigen Plane Waves}
\author{Ismo V.~Lindell and Ari~Sihvola\\ {\tt ismo.lindell@aalto.fi\quad ari.sihvola@aalto.fi}\\
{\it Department of Radio Science and Engineering}\\ {\it Aalto University, School of Electrical Engineering}\\
{\it Espoo, Finland}}

\maketitle
\begin{abstract}
It is known that the two eigen plane waves incident to the generalized soft-and-hard/DB (GSHDB) boundary are reflected as from the PEC or PMC boundary, i.e., with reflection coefficients $-1$ or $+1$, for any angle of incidence. The present paper discusses a more general class of boundaries by requiring that the reflection coefficients $R_+$ and $R_-$, corresponding to the two eigen plane waves, have opposite values, $R_\pm=\pm R$ with $R$ independent of the angle of incidence. It turns out that, there are two possibilities, $R=1$ for the class of GSHDB boundaries, and $R=j$ for another class, extending that of the perfect electromagnetic conductor (PEMC) boundaries. Matched waves at, and plane-waves reflected from, boundaries of the latter class are studied in the paper.
\end{abstract}

\section{Introduction}

Boundary conditions are fundamental building blocks in problems of mathematical physics, including electromagnetics. For given sources in a given space, Maxwell equations require additional constraints on the boundaries of the domain that force uniqueness and existence of the solution. These conditions between the electric and magnetic fields and fluxes at the boundary can in principle be rather general \cite{GBC}; however, in finite real-world problems, the boundaries often are characterized by approximate surface conditions \cite{Hoppe,Senior}. 

Examples of such boundary conditions are perfectly electrically conducting (PEC) surfaces, artificial magnetic conductor and high-impedance surfaces \cite{Sievenpiper1,Sievenpiper2}, and anisotropic soft-and-hard surfaces \cite{Kildal}. But  the possibilities to engineer the electromagnetic response of complex boundaries have a great variety. Thin two-dimensional structures which are capable of modifying the reflection and transmission of the incident wave in multiple manners have been recently designed and fabricated, and the label {\it metasurface} has appeared in the electromagnetics literature to cover such advanced surfaces \cite{Holloway}. Generalizing the traditional frequency-selective surfaces and transmit/reflect arrays \cite{Munk,Maci}, these metasurfaces can control the amplitude and phase of the wavefront in various ways \cite{Grbic}, the effects can be nonreciprocal \cite{Caloz}, non-local, or even active \cite{Kusti}. In the optical range of the electromagnetic spectrum, these developments within the metasurface paradigm are known by labels like planar photonics \cite{Kildishev} and flat optics \cite{Yu}.

It is the purpose of the present paper to investigate still uncharted classes of complex electromagnetic boundaries and surfaces. The treatment is purely theoretical, concentrating on certain aspects of electromagnetic boundary conditions beyond those in the book \cite{GBC} by these authors. Theoretical predictions of unconventional boundaries, like, as example,  the perfect electromagnetic conductor PEMC \cite{PEMC} have found experimental realizations \cite{Shahvarpour}, and found practical applications in the manipulation of properties of electromagnetic waves.  The novel class of boundary conditions introduced in the present paper may be of use in the future design of  novel metaboundaries.

\section{General Boundary Conditions}

A typical electromagnetic problem involves solving the Maxwell equations in a region bounded by a surface with imposed boundary conditions. The most general (linear and local) boundary conditions (GBC) can be shown to take the form \cite{GBC,PIER2016,AP2017},
\ea \A_1 c\#n\.\#B + \frac{\B_1}{\E_o}\#n\.\#D + \#a_{1t}\.\#E + \h_o\#b_{1t}\.\#H &=& 0, \l{BC1}\\
\A_2 c\#n\.\#B + \frac{\B_2}{\E_o}\#n\.\#D + \#a_{2t}\.\#E + \h_o\#b_{2t}\.\#H &=& 0, \l{BC2}\fa
with
\e c = 1/\sqrt{\M_o\E_o},\ \ \h_o=\sqrt{\h_o/\E_o}. \f
The unit vector $\#n$ is normal, and the vectors $\#a_{1t}\cdots \#b_{2t}$ are tangential, to the boundary surface. They are dimensionless and so are the scalars $\A_1,\A_2,\B_1,\B_2$. Because one of the scalar coefficients in each equation can be chosen, the number of free parameters defining the general boundary conditions is 10. Realization of a boundary defined by general conditions of the form \r{BC1} and \r{BC2}, in terms of the interface of a bianisotropic medium, has been discussed in \cite{GBC}, Section 5.14.

In the present paper we assume for simplicity that the boundary surface $\#n\.\#r=0$ is planar, as defined by a constant unit vector $\#n$. Also, we assume that the medium is isotropic with parameters $\E_o$ and $\M_o$, in which case the boundary conditions \r{BC1} and \r{BC2} can be rewritten in the simpler equivalent form \cite{GBC,AP2017}
\ea \#a_1\.\#E + \#b_1\.\h_o\#H &=&0, \l{bc1} \\
     \#a_2\.\#E + \#b_2\.\h_o\#H &=& 0, \l{bc2} \fa
where the four vectors $\#a_1\cdots\#b_2$ may contain components normal and tangential to the boundary,
\e \#a_i= a_{in}\#n + \#a_{it},\ \ \ \#b_i= b_{in}\#n + \#b_{it},\ \ \ i=1,2. \f
It is assumed that \r{bc1} and \r{bc2} represent two linearly independent conditions, because, otherwise, they do not define a unique boundary-value problem. 

\subsection{Special Cases}

The boundary conditions \r{bc1} and \r{bc2} contain a lot of familiar special cases starting from the perfect electric conductor (PEC), defined by
\e \#a_{1t}\.\#E = \#a_{2t}\.\#E=0,\ \ \ \ \#a_{1t}\x\#a_{2t}\not=0, \l{PEC} \f
and the perfect magnetic conductor (PMC), defined by
\e \#b_{1t}\.\#H=\#b_{2t}\.\#H=0,\ \ \ \ \#b_{1t}\x\#b_{2t}\not=0. \l{PMC}\f
A straightforward generalization of these two conditions is that of the perfect electromagnetic conductor (PEMC) \cite{PEMC,AP05,GPEMC},
\e \#b_{1t}\.(\#H + M\#E)= \#b_{2t}\.(\#H+ M\#E)=0, \l{PEMC} \f
or, in vector form,
\e \#n\x(\#H+ M\#E)=0, \l{PEMC1}\f
where $M$ is the PEMC admittance.

As other special cases of the GBC conditions \r{BC1} and \r{BC2} we may add
\begin{itemize}
\item The DB conditions \cite{DB},  
\e \#n\.\#D=\E_o\#n\.\#E=0,\ \ \ \ \#n\.\#B=\M_o\#n\.\#H=0. \l{DB} \f
\item The soft-and-hard (SH) conditions \cite{SHS1},
\e \#a_t\.\#E=0,\ \ \ \ \#a_t\.\#H=0. \l{SH} \f
\item The generalized soft-and-hard (GSH) conditions \cite{GSH},  
\e \#a_t\.\#E=0,\ \ \ \ \#b_t\.\#H=0. \l{GSH}\f
\item In the EH boundary conditions \cite{GBC},
\e \#a\.\#E=0,\ \ \ \ \#b\.\#H=0, \l{EH} \f
the vectors $\#a$ and $\#b$ are not restricted, whence \r{EH} are more general than  \r{GSH}.
\item The soft-and hard/DB (SHDB) conditions \cite{SHDB},
\e \#a_t\.\#E+ \A \#n\.\h_o\#H =0,\ \ \ \ \A \#n\.\#E - \#a_t\.\h_o\#H=0, \l{SHDB}\f
contain SH and DB conditions \r{SH}, \r{DB} as special cases. 
\item The generalized soft-and-hard/DB (GSHDB) conditions \cite{GSHDB},  
\e \#a_t\.\#E+  \A \#n\.\h_o\#H =0,\ \ \ \ \B \#n\.\#E +\#b_t\.\h_o\#H=0, \l{GSHDB}\f
represent a further generalization of \r{SHDB}. 
\item The impedance conditions  
\e \#a_{1t}\.\#E + \#b_{1t}\.\h_o\#H =0,\ \ \ \ \#a_{2t}\.\#E+ \#b_{2t}\.\h_o\#H=0. \l{imp}\f
\end{itemize}

\subsection{Purpose of this Study}

In a  previous study \cite{GSHDB}, it has been shown that a boundary defined by the GSHDB conditions \r{GSHDB} has the following property: any plane wave incident from an isotropic half space to the boundary can be decomposed in two eigen plane waves, one of which is reflected as from the PEC boundary, and, the other one, as from the PMC boundary. Since the GSHDB conditions \r{GSHDB} contain \r{DB} -- \r{GSH} and \r{SHDB} as special cases, the same property is shared by all the corresponding boundaries. The converse property was shown in \cite{PIER2016,GBC}: if the PEC/PMC decomposition property is required to be valid for the eigenwaves associated to the GBC conditions \r{bc1}, \r{bc2} (reflection coefficient either $+1$ or $-1$), the boundary must actually satisfy the GSHDB conditions \r{GSHDB}. Thus, the generalized soft-and-hard/DB boundary is the most general boundary with the PEC/PMC equivalence property.

It is the purpose of the present paper to extend this theory by requiring that the reflection coefficients $R_+,R_-$ corresponding to the two eigen plane waves satisfy $R_\pm=\pm R$, where $R$ is independent of the angle of incidence ($\#k$ vector of the incident wave). For the special case of the GSHDB boundary we have $R=1$.

\section{Reflection of Eigen Plane Waves}

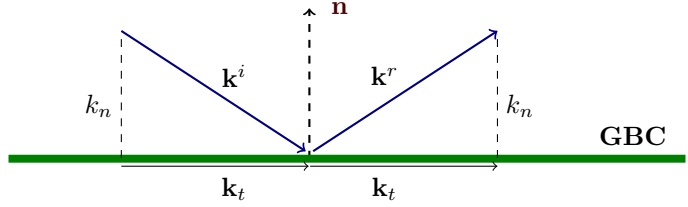
\begin{figure}[htbp]
	\centering
	\begin{tikzpicture}[scale=1.0]
		\draw[thick,dashed][->] (0,0) -- (0,2);
		\draw[green!50!black, line width=3pt] (-4,0) -- (5,0);
		\node[black] at (4.3,.3) {\bf{GBC}};
		\node at (-1,1.1) {$\mathbf{k}^i$};
		\node at (1,1.1) {$\mathbf{k}^r$};
		\node at (-1,-.4) {$\mathbf{k}_t$};
		\node at (1,-.4) {$\mathbf{k}_t$};
		\node at (-2.8,.7) {$k_n$};
		\node at (2.8,.7) {$k_n$};
		\node[red!30!black] at (.4,2) {$\mathbf{n}$};
		\draw[dashed] (-2.5,.01)--(-2.5,1.65);
		\draw[dashed] (2.5,.01)--(2.5,1.6);
		\draw [->] (.01,-.1)--(2.5,-.1);
		\draw [->] (-2.5,-.1)--(-.01,-.1);
		\draw[blue!50!black,thick] [->] (-2.5,1.7)--(-.05,7/70);
		\draw[blue!50!black,thick] [->] (.05,7/70)--(2.5,1.7);
	\end{tikzpicture}
	\caption{Plane wave incident on a surface characterized by a general boundary condition GBC. The incident and reflected wave vectors $\mathbf{k}^i,\mathbf{k}^r$ are decomposed into their tangential and normal components.}
	\label{fig:geometria}
\end{figure}

Let us consider time-harmonic plane waves satisfying the plane-wave equations
\ea \#k\x\#E &=& k_o\h_o\#H, \l{kxE}\\
\#k\x\h_o\#H &=& -k_o\#E. \l{kxH}\fa
%\e \#k\x\#E = k_o\M_o\#H,\ \ \ \#k\x\h_o\#H = -k_o\#E, \l{plane1} \f
and denote fields incident to, and reflecting from, the boundary surface, respectively by
\e \#E^i(\#r)= \#E^i\exp(-j\#k^i\.\#r),\ \ \ \ \#E^r(\#r)= \#E^r\exp(-j\#k^r\.\#r), \f
with (see Figure \ref{fig:geometria})
\e \#k^i = \#k_t -k_n\#n,\ \ \ \ \#k^r=\#k_t + k_n\#n,\f
\e \#k^i\.\#k^i=\#k^r\.\#k^r=k_o^2=\o^2\M_o\E_o. \f

Eigen plane waves are defined by requiring that the tangential components of the electric fields at the boundary satisfy a relation of the form
\e \#E_t^r=R\#E_t^i, \l{ERE}\f
i.e., that the incident and reflected fields tangential to the boundary have similar polarization and are related by of a scalar factor $R$, the reflection coefficient.

Polarizations and reflection coefficients of the eigen plane waves depend on the vector parameters of the GBC boundary conditions \r{bc1}, \r{bc2}. Applying a previous analysis \cite{GBC,PIER2016}, the reflection coefficient can be shown to satisfy the condition
\e \#n\.((1+R)\#a_{1t}' - (1-R)\#b_{1t}'\.\=J_t)\x((1+R)\#a_{2t}' -(1-R)\#b_{2t}'\.\=J_t)=0, \l{R+}\f
where we denote
\ea \#a_{1t}'&=&\#a_{1t}+ (b_{1n}/k_o)\#n\x\#k_t, \l{a1'} \\
\#a_{2t}'&=&\#a_{2t}+ (b_{2n}/k_o)\#n\x\#k_t, \\
\#b_{1t}'&=&\#b_{1t}- (a_{1n}/k_o)\#n\x\#k_t, \\
\#b_{2t}'&=&\#b_{2t}- (a_{2n}/k_o)\#n\x\#k_t, \l{b2'}\fa
and
\e \=J_t = \frac{1}{k_ok_n}((\#n\x\#k)\#k_t + k_n^2\#n\x\=I_t). \l{Jt}\f
Here we assume $k_n\not=0$, which excludes the possibility of lateral eigenwaves, $\#k=\#k_t$. The dyadic $\=J_t$ is known to satisfy the properties \cite{GBC}
\e \=J_t\.\=J_t = -\=I_t = -\=I + \#n\#n, \f
and
\e (\#c_t\.\=J_t)\x(\#d_t\.\=J_t) = \#c_t\x\#d_t, \f
valid for any two vectors $\#c_t,\#d_t$ tangential to the boundary. The rule
\e k_ok_n\#b'_{it}\.\=J_t = (\#b_{it}\.\#n\x\#k- k_oa_{in})\#k_t -k_n^2\#n\x\#b_{it}, \l{bitJt}\f
for $i=1,2$, appears useful in the subsequent analysis.

In the general case, the solutions $R$ of the equation \r{R+} depend on the boundary parameters $\#a_1\cdots\#b_2$ and the tangential component of the wave vector, $\#k_t$.  Requiring that the solutions $R$ be independent of the wave vector sets conditions for the boundary parameter vectors. For example, for the GSHDB boundary, \r{R+} has the solutions $R=+1$ and $R=-1$ for any $\#k_t$ \cite{PIER2016}.

Equation \r{R+} is of the quadratic form
\e C_2R^2+ C_1 R + C_0 =0, \l{CR}\f
with
\ea C_2 &=& \#n\.(\#a_{1t}'+\#b_{1t}'\.\=J_t)\x(\#a_{2t}'+ \#b_{2t}'\.\=J_t) , \l{CC2}\\
C_1 &=& \#n\.(\#a_{1t}'+\#b_{1t}'\.\=J_t)\x(\#a_{2t}'- \#b_{2t}'\.\=J_t) + \nonumber\\
&+&\#n\.(\#a_{1t}'-\#b_{1t}'\.\=J_t)\x(\#a_{2t}'+ \#b_{2t}'\.\=J_t), \l{CC1}\\
C_0 &=& \#n\.(\#a_{1t}'-\#b_{1t}'\.\=J_t)\x(\#a_{2t}'- \#b_{2t}'\.\=J_t). \l{CC0}\fa
For $C_2\not=0$ the two solutions are obtained from
\e R_\pm = -\frac{C_1}{2C_2}\pm \sqrt{\frac{C_1^2}{4C_2^2}- \frac{C_0}{C_2}}. \l{Rpm}\f

The coefficients $C_0$ and $C_2$ are closely related. From \r{CC2} and \r{CC0} we can write
\e k_ok_nC_0 = A-B,\ \ \ \ k_ok_nC_2=A+B, \f
with
\ea A &=& k_ok_n\#n\.(\#a_{1t}'\x\#a_{2t}'+ \#b_{1t}'\x\#b_{2t}'), \l{A}\\
    B &=&  k_ok_n\#n\.(\#a_{1t}'\x(\#b_{2t}'\.\=J_t)- \#a_{2t}'\x(\#b_{1t}'\.\=J_t)). \l{B}\fa
Expanding these expressions as shown in the Appendix, the three coefficients take the form
\ea C_0 &=& \frac{-1}{k_ok_n}(k_o(\#a_1\x\#a_2 + \#b_1\x\#b_2)\.\#k^i+(\#b_1\#a_2-\#b_2\#a_1):(k_o^2\=I-\#k^i\#k^i)), \l{C0'}\\
   C_1 &=& 2\#n\.(\#a_1\x\#a_2 - \#b_1\x\#b_2) -\frac{2}{k_o}(\#a_1\x\#b_2 + \#b_1\x\#a_2)\.(\#n\x\#k_t),  \l{koC1}\\
    C_2 &=&  \frac{1}{k_ok_n}(k_o(\#a_1\x\#a_2 + \#b_1\x\#b_2)\.\#k^r+(\#b_1\#a_2-\#b_2\#a_1):(k_o^2\=I-\#k^r\#k^r)). \l{C2'} \fa
%Defining the function
%\e C(\#k) = \frac{1}{k_o\#n\.\#k}\le(k_o(\#a_1\x\#a_2 + \#b_1\x\#b_2)\.\#k+(\#b_1\#a_2-\#b_2\#a_1):(k_o^2\=I-\#k\#k)%\ri, \l{Ck} \f
%we can represent the coefficients $C_2$ and $C_0$ by
%\e C_2 = C(\#k^r),\ \ \ \ C_0= C(\#k^i).\l{C20} \f
As a special case, for the GSHDB boundary \r{GSHDB}, defined by $a_{1n}=b_{2n}=0$ and $\#a_{2t}=\#b_{1t}=0$, we have $C_1=0$ and $C_0=-C_2$, whence, from \r{Rpm}, $R_\pm=\pm 1$.

Let us now concentrate on finding the boundary conditions corresponding to the relation $R_+=-R_-$.

\section{Boundaries Defined by $R_+=-R_-$}

From \r{Rpm} we find that $R_+=-R_-$ is equivalent to the condition
\e C_1=0, \l{C10}\f
without any restriction on $C_0$ and $C_2$. From \r{koC1} we obtain the corresponding relation between the vectors $\#a_1\cdots\#b_2$,
\e  k_o\#n\.(\#a_1\x\#a_2 - \#b_1\x\#b_2) +(\#n\x(\#a_1\x\#b_2 + \#b_1\x\#a_2))\.\#k_t =0. \f
Requiring this to be valid for any angle of incidence, i.e., for any tangential vector $\#k_t$, leads to the two conditions
\ea \#n\.(\#a_1\x\#a_2 - \#b_1\x\#b_2)&=&0, \l{cond1a} \\
 \#n\x(\#a_1\x\#b_2- \#a_2\x\#b_1)&=&0. \l{cond2a}  \fa
They are respectively equivalent to
 \ea \#a_{1t}\x\#a_{2t}-\#b_{1t}\x\#b_{2t} &=&0, \l{cond1}\\
 b_{2n}\#a_{1t} - b_{1n}\#a_{2t} + a_{2n}\#b_{1t}- a_{1n}\#b_{2t}&=&0. \l{cond2}\fa
For the GSHDB boundary \r{GSHDB},  these are identically satisfied. 

With \r{C10}, the condition \r{CR} becomes 
\e C_2R^2+C_0=0, \l{R2}\f
whence
\e R_\pm = \pm\sqrt{-C_0/C_2}. \f

Requiring that the solution $R$ of \r{R2} be independent of $\#k_t$, \r{R2} can be split in the following set of equations relating the boundary parameters $\#a_1\cdots\#b_2$ and the reflection coefficient $R$:
\ea (R^2-1)(\#a_1\x\#a_2 + \#b_1\x\#b_2)\x\#n &=&0, \l{odd1x}\\
 (R^2+1)\#n\.(\#b_1\#a_2+ \#a_2\#b_1-\#b_2\#a_1-\#a_1\#b_2)\x\#n &=& 0, \l{odd2x}\\
% (R^2-1)(\#b_{1t}\.\#a_{2t}-\#b_{2t}\.\#a_{1t}) &=&0, \l{even1x} \\
 (R^2+1)\#n\.(\#a_1\x\#a_2)&=& 0, \l{even2x}\\
 (R^2-1)(b_{1n}a_{2n}-b_{2n}a_{1n})&=&0, \l{even32x}\\
 (R^2-1)(\#b_{1t}\#a_{2t} + \#a_{2t}\#b_{1t}-\#b_{2t}\#a_{1t}- \#a_{1t}\#b_{2t})&=&0. \l{even31'dx}\fa
Derivation of this set of equations is given in the Appendix.

Now we can separate three possible cases:
\begin{itemize}
\item $R^2=1$
\item $R^2=-1$
\item $R^2\not=1$ and $R^2\not=-1$.
\end{itemize}

In the last case, all expressions in \r{odd1x} -- \r{even31'dx} multiplying $(R^2-1)$ or $(R^2+1)$ must vanish. Actually, this corresponds to setting $C_0=C_1=C_2=0$ in \r{CR}, which means that $R$ may have any value. Ignoring this, we have only two possibilities, either $R^2=1$, or $R^2=-1$. Each of these defines a class of boundaries through certain conditions for the vector parameters $\#a_1\cdots\#b_2$ arising from  \r{odd1x} -- \r{even31'dx}. Since it has been previously shown \cite{GBC,PIER2016} that the case $R^2=1$ corresponds to the class of GSHDB boundaries defined by \r{GSHDB}, let us concentrate on the case $R^2=-1$, which yields a second possibility for the class of boundaries satisfying $R_+=-R_-$. In this case, the reflection coefficients for the two eigenwaves are $+j$ and $-j$.

\section{Boundaries Defined by $R_+=-R_-=j$}

From \r{odd1x}, \r{even32x} and \r{even31'dx} we obtain conditions for the boundary corresponding to the case $R^2=-1$:
\ea a_{1n}\#a_{2t} -a_{2n}\#a_{1t} + b_{1n}\#b_{2t}- b_{2n}\#b_{1t} &=&0, \l{odd1x'}\\
 b_{1n}a_{2n}-b_{2n}a_{1n}&=&0. \l{even32x'}\\
\#b_{1t}\#a_{2t} + \#a_{2t}\#b_{1t}-\#b_{2t}\#a_{1t}- \#a_{1t}\#b_{2t}&=&0, \l{even31'dx'}\fa
which, together with \r{cond1} and \r{cond2}, define the present class of boundaries.

It has been shown that the case $R^2=1$ corresponds to the class of GSHDB boundaries, for which the vector parameters satisfy $\#n\.\#a_1\x\#a_2=0$ and $\#n\.\#b_1\x\#b_2=0$, as is seen from \r{GSHDB}. Similarly, from \r{cond1} and \r{even2x} we can conclude that the case $R^2=-1$ corresponds to $\#n\.\#a_1\x\#a_2\not=0$ and $\#n\.\#b_1\x\#b_2\not=0$. Because of this, we can assume that the tangential vectors $\#b_{1t},\#b_{2t}$ and $\#a_{1t},\#a_{2t}$ form two linearly independent pairs, whence we can expand
\e \am \#b_{1t}\\ \#b_{2t}\a = {\cal B} \am \#a_{1t} \\ \#a_{2t}\a, \l{b1tb2t}\f
for some matrix  
\e {\cal B} = \amm B_{11} & B_{12}\\ B_{21} & B_{22}\a, \f
defined by four scalars $B_{11}\cdots B_{22}$. The condition \r{cond1} requires
\e \det{\cal B} = B_{11}B_{22}- B_{12}B_{21}=1. \l{detB}\f
From \r{cond2} we obtain a relation between the normal components of the four vectors as
\e \am b_{1n}\\ b_{2n}\a = -\amm B_{22} & -B_{12}\\ -B_{21} & B_{11}\a \am a_{1n}\\ a_{2n}\a = -{\cal B}^{-1}\am a_{1n}\\ a_{2n}\a. \l{b1nb2n}\f
Combining these, we can write
\ea \#b_1 &=& B_{11}\#a_1 + B_{12}\#a_2 -\tr{\cal B}\ a_{1n}\#n, \l{b11}\\
\#b_2 &=& B_{21}\#a_1 + B_{22}\#a_2 - \tr{\cal B}\ a_{2n}\#n, \l{b22}\fa
with
\e \tr{\cal B} = B_{11}+ B_{22}. \f
%\ea \am \#b_1\\ \#b_2\a &=& {\cal B} \am \#a_{1t} \\ \#a_{2t}\a -{\cal B}^{-1} \#n\am a_{1n}\\ a_{2n}\a\\
%&=& {\cal B} \am \#a_{1} \\ \#a_{2}\a -({\cal B}+{\cal B}^{-1}) \#n\am a_{1n}\\ a_{2n}\a\\
%&=& ({\cal B}\=I - \tr{\cal B}\ \#n\#n)\.\am \#a_1\\ \#a_2\a. \l{b1b2}\fa
Applying \r{b11} and \r{b22},  the boundary conditions \r{bc1} and \r{bc2} take the form
\ea \#a_1\.\#E  + (B_{11}\#a_1 + B_{12}\#a_2- \tr{\cal B}\ a_{1n}\#n)\.\h_o\#H &=& 0, \l{a1E}\\
\#a_2\.\#E + (B_{21}\#a_1 + B_{22}\#a_2 - \tr{\cal B}\ a_{2n}\#n)\.\h_o\#H &=& 0, \l{a2E}\fa
for some vectors $\#a_1,\#a_2$ and a matrix ${\cal B}$ restricted by \r{detB}. 

Applying \r{b1tb2t}, \r{even31'dx'} becomes
\e -2B_{21}\#a_{1t}\#a_{1t}+ (B_{11}-B_{22})(\#a_{1t}\#a_{2t}+ \#a_{2t}\#a_{1t}) + 2B_{12}\#a_{2t}\#a_{2t}=0, \f
which for $\#a_{1t}\x\#a_{2t}\not=0$ yields
\e B_{12}=B_{21}=0,\ \ \ B_{11}=B_{22}=B, \ \ \ \tr{\cal B}=2B,\f
for some scalar $B$. From \r{detB} we obtain
\e \det{\cal B}= B_{11}B_{22}-B_{12}B_{21} = B^2=1. \f
with two possible solutions $B=+1$ and $B=-1$.

From \r{b11} and \r{b22} we obtain the corresponding possibilities,
\e \#b_1=\#a_{1t}-\#n a_{1n},\ \ \ \#b_2 = \#a_{2t}- \#n a_{2n}, \l{b12B+}\f
and
\e \#b_1=-\#a_{1t}+\#n a_{1n},\ \ \ \#b_2 = -\#a_{2t}+ \#n a_{2n}. \l{b12B-}\f
\r{odd1x'} and \r{even32x'} are now satisfied for any given vectors $\#a_1$ and $\#a_2$. Substituting \r{b12B+} or \r{b12B-}, the conditions \r{cond1a}, \r{cond2a} and \r{odd1x} -- \r{even31'dx} are also satisfied. 

Thus, the above analysis offers two possibilities for the boundary conditions corresponding to the case $R^2=-1$,  
\ea \#a_{1t}\.(\#E +\h_o\#H) + a_{1n}\#n\.(\#E-\h_o\#H) &=& 0, \l{a1EB+}\\
\#a_{2t}\.(\#E +\h_o\#H) + a_{2n}\#n\.(\#E-\h_o\#H)  &=& 0, \l{a2EB+} \fa
and
\ea \#a_{1t}\.(\#E -\h_o\#H) + a_{1n}\#n\.(\#E+\h_o\#H) &=& 0, \l{a1EB-}\\
\#a_{2t}\.(\#E -\h_o\#H) + a_{2n}\#n\.(\#E+\h_o\#H)  &=& 0. \l{a2EB-} \fa
The conditions \r{a1EB+}, \r{a2EB+} can be expressed in vector form as
\e \#n\x(\#E+\h_o\#H) + \#p_t\#n\.(\#E-\h_o\#H)=0, \l{EPEMC+} \f
and the conditions \r{a1EB-}, \r{a2EB-} as
\e \#n\x(\#E-\h_o\#H) + \#p_t\#n\.(\#E+\h_o\#H)=0, \l{EPEMC-} \f
with
\ea \#p_t &=& \frac{1}{A_n}(\#a_1\x\#a_2)\x\#n,
= \frac{1}{A_n}(a_{1n}\#a_{2t}- a_{2n}\#a_{1t}),  \l{pt} \\
A_n &=& \#n\.\#a_1\x\#a_2. \l{An} \fa
In the present case $R^2=-1$ we have assumed $A_n\not=0$.

In the special case of impedance boundaries, the boundary parameters satisfy $a_{1n}=a_{2n}=0$, which corresponds to $\#p_t=0$. In this case, the condition \r{EPEMC+} becomes
\e \#n\x(\#E +\h_o\#H) = 0, \l{EPEMC++}\f
and, the condition \r{EPEMC-} becomes
\e \#n\x(\#E -\h_o\#H) = 0. \l{EPEMC--}\f
\r{EPEMC++} and \r{EPEMC--} equal special cases of the PEMC boundary conditions \r{PEMC}, with admittances $M=1/\h_o$  and $M=-1/\h_o$, respectively. The reflection coefficients for the two eigenwaves at the general PEMC boundary are known to have the form \cite{GBC}, Eq. (2.38),
\e R_\pm = \frac{1\pm jM\h_o}{1\mp jM\h_o}, \f
which satisfies $R_\pm^2=-1$ for both $M\h_o=1$ and $M\h_o=-1$.

One can easily verify that, for both of the boundary conditions \r{EPEMC+} and \r{EPEMC-}, the two solutions for the reflection coefficient equation \r{CR} satisfy both $R_+=-R_-$ and $R_\pm^2=-1$. The details are given in the Appendix.

In conclusion, requiring that the reflection coefficients $R_+$ and $R_-$ of the two eigenwaves satisfy the condition $R_+=-R_-=R$, with $R^2=-1$ for any angle of incidence, the boundary is defined either by the conditions \r{EPEMC+} or \r{EPEMC-}.

\section{Extended PEMC Boundaries}

The conditions \r{EPEMC+} and  \r{EPEMC-} are special cases of the vector condition
\e \#n\x(M\#E + \#H) + \#p_t\#n\.(M\#E-\#H) =0, \l{EPEMC} \f
defining what can be called the class of Extended PEMC (EPEMC) boundaries. The class depends on two parameters, a scalar $M$ and a vector $\#p_t$. The boundary vector parameters corresponding to \r{EPEMC} are 
\e \#b_1=(\#a_{1t}-\#n a_{1n})/M\h_o,\ \ \ \#b_2=(\#a_{2t}- \#n a_{2n})/M\h_o. \l{b1b2}\f
For $\#p_t=0$, the condition of the EPEMC boundary \r{EPEMC} is reduced to that of the PEMC boundary \r{PEMC}, in which case the parameter $M$ equals the admittance of the PEMC boundary. 

Because \r{EPEMC+} and \r{EPEMC-} are special cases of \r{EPEMC}, respectively corresponding to $M=1/\h_o$ and $M=-1/\h_o$, they can be called conditions of the EPEMC$_+$ and EPEMC$_-$ boundaries, respectively. 
For $\#p_t=0$ both of them are reduced to the corresponding special PEMC boundaries. 

It has been previously shown that a plane wave incident normally to the PEMC boundary is reflected cross polarized, i.e., satisfying $\#E_t^r\.\#E_t^i=0$, exactly when the PEMC admittance satisfies either $M\h_o=+1$ or $=-1$ \cite{GBC}. It will turn out that the same property is valid for the EPEMC boundary.

\subsection{Matched Waves}

By definition, a plane wave is matched to a boundary when the boundary conditions are identically satisfied for the single plane wave, i.e., there is no reflected wave \cite{GBC}. Applying the plane-wave equations \r{kxE} and \r{kxH}, the conditions of the  EPEMC$_+$ and EPEMC$_-$ boundaries, \r{EPEMC+} and \r{EPEMC-}, can be respectively  expanded as 
\ea k_o\#n\x(\#E+\h_o\#H) &=& - \#p_t\#n\.k_o(\#E-\h_o\#H)\nonumber\\
&=& -\#p_t\#k_t\.(\#n\x(\#E+\h_o\#H)), \fa
\ea k_o\#n\x(\#E-\h_o\#H) &=& - \#p_t\#n\.k_o(\#E+\h_o\#H)\nonumber\\
&=&\#p_t\#k_t\.(\#n\x(\#E-\h_o\#H)). \fa
Thus, the conditions for the matched waves at the respective EPEMC$_+$ and EPEMC$_-$ boundaries become
\ea (k_o\=I_t + \#p_t\#k_t)\.(\#n\x(\#E+ \h_o\#H)) &=& 0, \l{match+}\\
(k_o\=I_t - \#p_t\#k_t)\.(\#n\x(\#E- \h_o\#H)) &=& 0. \l{match-}\fa
Let us consider these two cases separately.

\subsubsection {EPEMC$_+$ Boundary}  

From \r{match+} we have two possibilities: either
\e \#n\x(\#E+ \h_o\#H)=0, \l{match+1} \f
or the two-dimensional dyadic in \r{match+} has no inverse. 

In the latter case, the two-dimensional determinant of the dyadic must be zero \cite{GBC,Methods}, whence
\e \frac{1}{k_o}\tr(k_o\=I_t + \#p_t\#k_t)^{(2)} = k_o + \#p_t\.\#k_t =0, \l{disp1} \f
which is the dispersion equation of the matched wave. Because the dyadic can be expressed as
\e k_o\=I_t + \#p_t\#k_t = \#k_t\x(\#p_t\x\=I_t)= -(\#n\x\#k_t)(\#n\x\#p_t), \f
from \r{match+} the fields of the matched wave must satisfy 
\e (\#n\x\#p_t)\.(\#n\x(\#E+\h_o\#H) = \#p_t\.(\#E+\h_o\#H)=0. \f
Applying the orthogonality $\#k\.(\#E+\h_o\#H)=0$, the polarization of the matched wave corresponding to the dispersion equation \r{disp1} is
\e \#E+\h_o\#H\sim \#k\x\#p_t. \f

Considering the case \r{match+1}, from \r{EPEMC+} we obtain
\e \#n\.(\#E-\h_o\#H)=0. \l{nE-hH}\f
Applying \r{kxE} and \r{kxH} we can expand 
\ea k_o\#n\x(\#E+ \h_o\#H)&=& \#n\x(-\#k\x\h_o\#H + \#k\x\#E) \nonumber\\
&=& \#k\#n\.(\#E -\h_o\#H)- (\#n\.\#k)(\#E-\h_o\#H)\nonumber\\
&=& - (\#n\.\#k)(\#E-\h_o\#H)=0. \l{disp2}\fa
Thus, the second possible matched wave is governed by the dispersion equation 
\e \#n\.\#k=0, \l{nk0}\f
which corresponds to a lateral wave. From \r{nE-hH} we obtain the polarization condition
\e \#E -\h_o\#H \sim \#n\x\#k. \f

\subsubsection{EPEMC$_-$ Boundary}

This case can be handled similarly. From \r{match-}, for $\#n\x(\#E-\h_o\#H)\not=0$, the dispersion equation becomes
\e k_o-\#p_t\.\#k_t=0. \f
Because the fields must satisfy
\e \#p_t\.(\#E-\h_o\#H)=0, \f
the polarization condition becomes
\e \#E-\h_o\#H \sim \#k\x\#p_t. \f
For the other possibility  $\#n\x(\#E-\h_o\#H)=0$, the matched wave is again a lateral wave with dispersion equation \r{nk0}. From \r{EPEMC-} we now obtain
\e \#n\.(\#E+\h_o\#H)=0, \f
whence the polarization of the wave is
\e \#E+\h_o\#H\sim \#n\x\#k. \f

\subsection{Reflection of Plane Waves}

Let us finally consider plane-wave reflection from the EPEMC$_+$ and EPEMC$_-$ boundaries, defined by \r{EPEMC+} and \r{EPEMC-}.

Applying \r{kxE} and \r{kxH}, the sum of incident and reflected fields, at the  EPEMC$_+$ boundary can be shown to satisfy

\ea && k_o\#n\x(\#E^i+\h_o\#H^i +\#E^r+\h_o\#H^r) \nonumber\\
&=& - \#p_t\#n\.k_o(\#E^i-\h_o\#H^i+\#E^r-\h_o\#H^r)) \nonumber\\
&=& \#p_t\#n\.(\#k^i\x(\#E^i+\h_o\#H^i)+\#k^r\x(\#E^r+\h_o\#H^r))\nonumber\\
&=& -\#p_t\#k_t\.\#n\x(\#E^i+\h_o\#H^i + \#E^r+\h_o\#H^r).  \fa
The result can be written as
\e (k_o\=I_t + \#p_t\#k_t)\.\#n\x(\#E^i+\h_o\#H^i+\#E^r+\h_o\#H^r) =0. \l{EiEr+}\f
Omitting the case when $\#k$ satisfies the dispersion equation \r{disp1} of the matched wave, the dyadic in \r{EiEr+} has a two-dimensional inverse. In this case, the fields at the boundary must satisfy
\e \#n\x(\#E^i+\#E^r+\h_o(\#H^i+\#H^r)) =0, \l{condPEMC+}\f
for any vector $\#p_t$. Actually, \r{condPEMC+} equals the boundary condition of the PEMC boundary \r{PEMC1}, for $M=1/\h_o$. 

The EPEMC$_-$ boundary can be handled similarly by starting from the condition \r{EPEMC-}, leading to
\e (k_o\=I_t - \#p_t\#k_t)\.\#n\x(\#E^i-\h_o\#H^i+\#E^r-\h_o\#H^r) =0, \l{EiEr-}\f
whence the result \r{condPEMC+} will be replaced by 
\e \#n\x(\#E^i+\#E^r-\h_o(\#H^i+\#H^r)) =0, \l{condPEMC-}\f
for any vector $\#p_t$. Again, this equals the boundary condition of the PEMC boundary \r{PEMC1}, for $M=-1/\h_o$.

To summarize, the class of boundaries defined by the property $R^2=-1$ consists of two subclasses, EPEMC$_+$ and EPEMC$_-$, which are equivalent in reflection to two subclasses of PEMC boundaries, respectively defined by $M=1/\h_o$ and $M=-1/\h_o$, for any vector $\#p_t$.

\begin{figure}
\begin{center}
{
\includegraphics[width=11cm]{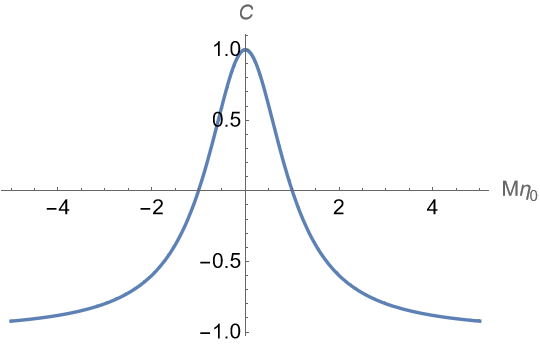}
}
\caption{\label{fig:copol.png}
Visualization of the co-polarization quantity $C$ of \r{C}, as a function of the parameter $M\h_o$, for a normally incident plane wave reflecting from an EPEMC boundary defined by \r{EPEMC}. The reflected field is co-polarized for $M\h_o=0$ (PMC) and $M\h_o\ra\pm\oo$ (PEC) while, for $M\h_o=1$ (EPEMC$_+$) and $M\h_o=-1$ (EPEMC$_-$),  the reflected field is cross polarized, $C=0$.}
\end{center}
\end{figure}

As a example, let us consider a plane wave with normal incidence, $\#k^r=-\#k^i=\#n k_o$, reflecting from an EPEMC boundary. The fields are tangential to the boundary: 
$\#n\.\#E^i=\#n\.\#E^r=0$, $\#n\.\#H^i=\#n\.\#H^r=0$. From \r{EPEMC}, the reflected field can be expressed as
\e \#E^r = -\frac{1}{M^2\h_o^2+1}\le (M^2\h_o^2-1)\=I_t + 2M\h_o\#n\x\=I\ri\.\#E^i, \l{ErC}\f
whence
\e \#E^r\.\#E^i = C(\#E^i\.\#E^i), \f
where $C$, defined by
\e C = -\frac{M^2\h_o^2-1}{M^2\h_o^2+1}, \l{C}\f
is a measure of copolarization in the reflected field. This function is depicted in Figure \ref{fig:copol.png}.

The reflected field appears cross polarized when $\#E^r\.\#E^i=0$, or $C=0$. This happens for $M\h_o=\pm1$, i.e., when the EPEMC boundary is either EPEMC$_+$ or EPEMC$_-$. The same property is known to exist for PEMC boundaries with $M=1/\h_o$ or $M=-1/\h_o$ (\cite{GBC}, Section 2.4).  For $M=0$ (PMC, $C=1$) and $M\ra\pm\oo$ (PEC, $C=-1$), the reflected field is totally copolarized. Note that, for a circularly polarized incident wave, satisfying $\#E^i\.\#E^i=0$, from \r{ErC} we have $\#E^r\.\#E^i=0$ for any $M\not= \pm j/\h_o$. 

\section{Conclusion}The problem considered in this paper has been to generalize the property of the class of electromagnetic boundaries known as that of generalized soft-and-hard/DB (GSHDB) boundaries. The class of boundaries is known to possess the unique property that the two eigen plane-waves associated to the boundary are reflected with reflection coefficients $R_+=1$ and $R_-=-1$ for any angle of incidence. Such a property has been called PEC/PMC equivalence, because the two eigenwaves are reflected as from a PEC or a PMC boundary. In the present case, the possibility of a class of boundaries with a more general property was studied by requesting that the reflection coefficients of the two eigenwaves satisfy a condition of the form $R_\pm=\pm R$, where $R$ is independent of the angle of incidence. 

It was shown that, under this condition, $R$ must be either $1$ or $j$. While the former case corresponds to the class of GSHDB boundaries, the latter case was shown to correspond to another class of boundaries defined by two possible sets of boundary conditions, \r{EPEMC+} and \r{EPEMC-}. Both of these were shown to be special cases of a more general class of Extended Perfect Electromagnetic Conductor (EPEMC) boundaries defined by conditions of the form \r{EPEMC}, depending on a scalar $M$ and a vector $\#p_t$ tangential to the boundary surface. The two special cases of \r{EPEMC} were respectively called EPEMC$_+$ and EPEMC$_-$ boundaries as they correspond to the special parameter values $M=1/\h_o$ and $M=-1/\h_o$, respectively. For $\#p_t=0$, the EPEMC boundary is reduced to the PEMC boundary with $M$ equal to the PEMC admittance. Matched waves and plane-wave reflection properties were studied for both EPEMC$_+$ and EPEMC$_-$ classes of boundaries. While the matched waves depend on the $\#p_t$ vector, reflection of the eigenwaves does not. Actually, in reflection, the EPEMC$_+$ and EPEMC$_-$ classes of boundaries are equivalent to the corresponding classes of PEMC boundaries defined by the respective admittance values $M=1/\h_o$ and $M=-1/\h_o$, for any vector $\#p_t$.

\section{Appendix}

\subsection{Expanding $A$ and $B$}

Applying \r{bitJt}, we can expand the expressions \r{A} and \r{B} as
\ea A &=& k_ok_n\#n\.(\#a_1\x\#a_2 + \#b_1\x\#b_2)+ k_n\#n\.(\#b_2\#a_1-\#b_1\#a_2 + \#a_1\#b_2-\#a_2\#b_1)\.\#k_t \nonumber\\
&=& k_ok_n\#n\.(\#a_1\x\#a_2 + \#b_1\x\#b_2)- (\#b_1\#a_2-\#b_2\#a_1):k_n(\#n\#k_t + \#k_t\#n), \\
 B &=& (\#b_{1t}'\#a_{2t}'-\#b_{2t}'\#a_{1t}'):(\#n\x\#k)(\#n\x\#k)+ k_n^2(\#a_{2t}'\.\#b_{1t}'-\#a_{1t}'\.\#b_{2t}') \nonumber\\
&=& (\#b_{1t}'\#a_{2t}'-\#b_{2t}'\#a_{1t}'):(k_t^2\=I_t- \#k_t\#k_t + k_n^2\=I_t) \nonumber\\
&=& (\#b_{1t}'\#a_{2t}'-\#b_{2t}'\#a_{1t}'):(k_o^2\=I_t- \#k_t\#k_t).  \fa
Substituting
\ea \#b_{1t}'\#a_{2t}'-\#b_{2t}'\#a_{1t} &=& \#b_{1t}\#a_{2t}-\#b_{2t}\#a_{1t}\nonumber\\
&+& \frac{1}{k_o}\bigg((b_{2n}\#b_{1t}-b_{1n}\#b_{2t})\#n\x\#k -\#n\x\#k(a_{1n}\#a_{2t}-a_{2n}\#a_{1t})\bigg)\nonumber\\
&-& \frac{1}{k_o^2}(a_{1n}b_{2n}-a_{2n}b_{1n})(\#n\x\#k)(\#n\x\#k), \fa
we obtain
\ea B &=& (\#b_1\#a_2-\#b_2\#a_1):(k_o^2\=I_t- \#k_t\#k_t)\nonumber\\
&+& k_o(b_{2n}\#b_1-b_{1n}\#b_2 -a_{1n}\#a_2+a_{2n}\#a_1)\.(\#n\x\#k)- (a_{1n}b_{2n}-a_{2n}b_{1n})k_t^2, \nonumber\\
&=& (\#b_1\#a_2-\#b_2\#a_1):(k_o^2\=I- \#k_t\#k_t- k_n^2\#n\#n)+ k_o(\#a_1\x\#a_2+\#b_1\x\#b_2)\.\#k_t. \fa
These can be combined as
\ea A\pm B &=& \pm k_o(\#a_1\x\#a_2+\#b_1\x\#b_2)\.(\#k_t \pm k_n\#n)\nonumber\\
&&\pm (\#b_1\#a_2-\#b_2\#a_1):(k_o^2\=I- \#k_t\#k_t- k_n^2\#n\#n\mp k_n(\#n\#k_t + \#k_t\#n))\nonumber\\
&=& \pm k_o(\#a_1\x\#a_2+\#b_1\x\#b_2)\.(\#k_t \pm k_n\#n)\nonumber\\
&&\pm (\#b_1\#a_2-\#b_2\#a_1):(k_o^2\=I- (\#k_t \pm k_n\#n)(\#k_t \pm k_n\#n))  \fa
Thus, we can finally write
\e k_ok_nC_0 = -k_o(\#a_1\x\#a_2 + \#b_1\x\#b_2)\.\#k^i-(\#b_1\#a_2-\#b_2\#a_1):(k_o^2\=I-\#k^i\#k^i), \l{koknC0'}\f
and
\e    k_ok_nC_2 =  k_o(\#a_1\x\#a_2 + \#b_1\x\#b_2)\.\#k^r+(\#b_1\#a_2-\#b_2\#a_1):(k_o^2\=I-\#k^r\#k^r). \l{koknC2'} \f

\subsection{Expanding Condition \r{R2} }

Applying \r{C0'} and \r{C2'}, the condition \r{R2} can be expressed in compact form as
\e (R^2-1)(K_o+K_e) + (R^2+1)(L_o+L_e)=0, \l{KoKeLoLe}\f
with terms odd in $\#k_t$,
\ea K_o &=& k_o(\#a_1\x\#a_2 + \#b_1\x\#b_2)\.\#k_t \l{Ko}\\
L_o &=& -k_n\#n\.(\#b_1\#a_2+ \#a_2\#b_1-\#b_2\#a_1-\#a_1\#b_2)\.\#k_t, \l{Lo}\fa
and, terms even in $\#k_t$
\ea K_e&=& (\#b_1\#a_2-\#b_2\#a_1):(k_o^2\=I_t -\#k_t\#k_t +(\#k_t\.\#k_t)\#n\#n)\l{Ke} \\
L_e&=& 2k_n k_o\#n\.(\#a_1\x\#a_2) . \l{Le}\fa
For \r{KoKeLoLe} to be satisfied for any $\#k_t$, parts odd and even in $\#k_t$ must vanish separately. This leads to the respective conditions
\ea (R^2-1)K_o + (R^2+1)L_o &=&0, \l{odd}\\
 (R^2-1) K_e+ (R^2+1)L_e&=&0. \l{even} \fa

As a special case, for the GSHDB boundary \r{GSHDB} with $\#a_1=\#a_t$, $\#b_1=\A\#n$, $\#a_2=\B\#n$ and $\#b_2=\#b_t$, we find $L_o=0$ and $L_e=0$. Since $\#a_t$ and $\#b_t$ may be any tangential vectors and $\A,\B$ any scalars, $K_o$ and $K_e$ are nonzero. Thus, both \r{odd} and \r{even} yield $R^2=1$, or $R_\pm=\pm 1$, a result known from \cite{GBC,GSHDB}.

Let us consider the general case. For \r{odd} to be valid for any polarization of $\#k_t$, we must have  
\ea (R^2-1)k_o(\#a_1\x\#a_2 + \#b_1\x\#b_2)\x\#n &+& \nonumber\\
- (R^2+1)k_n\#n\.(\#b_1\#a_2+ \#a_2\#b_1-\#b_2\#a_1-\#a_1\#b_2))\x\#n &=& 0. \fa
Since only the second term depends on $\#k_t$ (through $k_n$), the two terms must vanish independently, whence we can write
\ea (R^2-1)(a_{1n}\#a_{2t} -a_{2n}\#a_{1t} + b_{1n}\#b_{2t}- b_{2n}\#b_{1t}) &=&0, \l{odd1}\\
 (R^2+1)(b_{1n}\#a_{2t}+ a_{2n}\#b_{1t}- b_{2n}\#a_{1t} -a_{1n}\#b_{2t}) &=& 0. \l{odd2}\fa
These two conditions correspond to the equation \r{odd}. Similarly, \r{even} can be split in three conditions, based on  different orders of magnitude in $\#k_t$, as
\ea (R^2-1)(\#b_1\#a_2-\#b_2\#a_1):\=I_t &=&0, \l{even1} \\
 (R^2+1)\#n\.(\#a_1\x\#a_2)&=& 0, \l{even2}\\
 (R^2-1)(\#b_1\#a_2-\#b_2\#a_1):(\#k_t\#k_t -(\#k_t\.\#k_t)\#n\#n)&=&0. \l{even3}\fa
The condition \r{even3} can be further split in two parts because one of the terms does not depend on the polarization of $\#k_t$,
\ea (R^2-1)(\#b_1\#a_2-\#b_2\#a_1):\#k_t\#k_t&=&0, \l{even31}\\
 (R^2-1)(b_{1n}a_{2n}-b_{2n}a_{1n})&=&0. \l{even32}\fa
Setting $\#k_t=\#k_t'+\#k_t''$ in \r{even31}, we obtain 
\e (R^2-1)(\#b_1\#a_2-\#b_2\#a_1):(\#k_t'\#k_t''+\#k_t''\#k_t')=0, \l{even31's}\f
which must be valid for any $\#k_t'$ and $\#k_t''$, whence \r{even31's} equals the dyadic condition
\e (R^2-1)(\#b_{1t}\#a_{2t} + \#a_{2t}\#b_{1t}-\#b_{2t}\#a_{1t}- \#a_{1t}\#b_{2t})=0. \l{even31'd}\f
Since the trace of \r{even31'd} covers the condition \r{even1}, the latter can be omitted.

In conclusion, the medium parameters $\#a_1\cdots\#b_2$ and the reflection coefficient $R$ must satisfy the conditions \r{odd1}, \r{odd2}, \r{even2}, \r{even32} and \r{even31'd}

\subsection{Expanding $C_1=0$}

Let us find the conditions for the boundary parameters $\#a_1\cdots\#b_2$ in the case when the coefficient $C_1$ vanishes for all wave vectors $\#k_t$. Because the second term of \r{koC1} changes sign in $\#k_t\ra\#k_t$, from $C_1=0$ we obtain the two conditions
\ea \#n\.(\#a_1\x\#a_2 - \#b_1\x\#b_2) &=& 0, \l{C11}\\
(\#a_1\x\#b_2 + \#b_1\x\#a_2)\x\#n&=&0. \l{C12}\fa
Assuming $\#n\.\#a_1\x\#a_2\not=0$, the tangential vectors $\#a_{1t},\#a_{2t}$ make a planar basis and we can expand
\e \am \#b_{1t}\\ \#b_{2t}\a = \amm B_{11}& B_{12}\\ B_{21}& B_{22}\a \am \#a_{1t}\\ \#a_{2t}\a. \l{b1tb2t''}\f
Inserting in \r{C11}, we obtain
\e B_{11}B_{22}-B_{12}B_{21}=1. \l{detB11}\f
Expanding \r{C12} as
\e a_{1n}\#b_{2t}- b_{2n}\#a_{1t} + b_{1n}\#a_{2t}- a_{2n}\#b_{1t}=0, \f
and applying \r{b1tb2t''}, with $\#a_{1t}\x\#a_{2t}\not=0$, after some steps we obtain the expansion
\e \am b_{1n}\\ b_{2n}\a = \amm -B_{22} & B_{12}\\ B_{21} & -B_{11}\a \am a_{1n}\\ a_{2n}\a = - \amm B_{11}& B_{12}\\ B_{21}& B_{22}\a^{-1}\am a_{1n}\\ a_{2n}\a. \l{b1nb2n'} \f
In conclusion, vectors $\#b_1$ and $\#b_2$, depending through \r{b1tb2t''} and \r{b1nb2n'} on any vectors $\#a_1$ and $\#a_2$ whose tangential components are linearly independent, in terms of any parameters $B_{11}\cdots B_{22}$ satisfying \r{detB11}, yield $C_1=0$ for any $\#k_t$.

\subsection{Verifying EPEMC$_+$ and EPEMC$_-$ boundary conditions}

Let us verify that the EPEMC$_+$ boundary conditions \r{a1EB+} and \r{a2EB+}, and the EPEMC$_-$ boundary conditions \r{a1EB-} and \r{a2EB-}, depending on the two vectors $\#a_1$ and $\#a_2$, both correspond to the property $R_\pm^2=-1$. 

Substituting \r{b12B+} and \r{b12B-} and expanding
\ea \#a_1\x\#a_2 &=& \#a_{1t}\x\#a_{2t} + \#n\x(a_{1n}\#a_{2t}- a_{2n}\#a_{1t}), \\
 \#b_1\x\#b_2 &=& \#a_{1t}\x\#a_{2t} -\#n\x(a_{1n}\#a_{2t}- a_{2n}\#a_{1t}), \fa
we obtain
\e \#n\.(\#a_1\x\#a_2- \#b_1\x\#b_2) = 0. \f
Similarly, we can expand
\e (\#a_1\x\#b_2+ \#b_1\x\#a_2)\x\#n = \pm(a_{1n}\#a_{2t}+a_{2n}\#a_{1t} -a_{1n}\#a_{2t}-a_{2n}\#a_{1t})=0. \f
Substituting these in \r{koC1} we obtain 
\e C_1=0, \f
which, from \r{Rpm}, ensures the validity of the condition $R_+=-R_-$.

Further, we can expand
\ea \#a_1\x\#a_2+ \#b_1\x\#b_2 &=& 2A_n\#n, \\
\#b_1\#a_2-\#b_2\#a_1 &=& \mp A_n(\#n\x\=I + \#n\#p_t+ \#p_t\#n), \fa
with the upper sign corresponding to EPEMC$_+$, and the lower sign to EPEMC$_-$, boundary conditions. 

Substituting these in \r{C0'} and \r{C2'} yields
\ea C_0 &=& \frac{-2\#n\.\#k^i}{k_ok_n}A_n(k_o \pm\#p_t\.\#k_t), \\
 C_2 &=& \frac{2\#n\.\#k^r}{k_ok_n}A_n(k_o \pm\#p_t\.\#k_t) = C_0, \l{C2=C0}\fa
whence, from \r{Rpm}, the reflection coefficients are $j$ and $-j$ for both of the EPEMC$_+$ and EPEMC$_-$ conditions.

\end{document}